\documentclass[conference]{IEEEtran}
\IEEEoverridecommandlockouts
\usepackage{cite}
\usepackage{amssymb,amsfonts}
\usepackage{amsmath}
\usepackage{algorithm}
\usepackage{algpseudocode}
\usepackage{graphicx}
\usepackage{textcomp}
\usepackage{xcolor}
\usepackage{fancyhdr}
\usepackage[normalem]{ulem}
\usepackage{algpseudocode}
\usepackage{subfigure}
\usepackage{cancel}
\def\BibTeX{{\rm B\kern-.05em{\sc i\kern-.025em b}\kern-.08em
    T\kern-.1667em\lower.7ex\hbox{E}\kern-.125emX}}

\newboolean{showcomments}
\setboolean{showcomments}{true}
\ifthenelse{\boolean{showcomments}}
{
}

\newcommand*{\affmark}[1][*]{\textsuperscript{#1}}

\begin{document}

\title{Intent-driven Intelligent Control and Orchestration in O-RAN Via Hierarchical Reinforcement Learning}

\author{\IEEEauthorblockN{Md~Arafat~Habib\affmark[1], Hao~Zhou\affmark[1], Pedro~Enrique~Iturria-Rivera\affmark[1], Medhat Elsayed\affmark[2], Majid Bavand\affmark[2], \\Raimundas Gaigalas\affmark[2], Yigit Ozcan\affmark[2] and  Melike Erol-Kantarci\affmark[1], \IEEEmembership{Senior Member,~IEEE}}
\IEEEauthorblockA{\affmark[1]\textit{School of Electrical Engineering and Computer Science, University of Ottawa, Ottawa, Canada}}  \affmark[2]\textit{Ericsson Inc., Ottawa, Canada}\\
Emails:\{mhabi050, hzhou098, pitur008, melike.erolkantarci\}@uottawa.ca, \\\{medhat.elsayed, majid.bavand, raimundas.gaigalas, yigit.ozcan\}@ericsson.com \vspace{-1em}}

\maketitle

\thispagestyle{fancy}   
\fancyhead{}                
\lhead{Accepted by the 20th IEEE International Conference on
Mobile Ad-Hoc and Smart Systems
(MASS 2023)\copyright2023 IEEE}
\cfoot{}
\renewcommand{\headrulewidth}{0pt} 

\begin{abstract}
rApps and xApps need to be controlled and orchestrated well in the open radio access network (O-RAN) so that they can deliver a guaranteed network performance in a complex multi-vendor environment.  
This paper proposes a novel intent-driven intelligent control and orchestration scheme based on hierarchical reinforcement learning (HRL). The proposed scheme can orchestrate multiple rApps or xApps according to the operator's intent of optimizing certain key performance indicators (KPIs), such as throughput, energy efficiency, and latency. Specifically, we propose a bi-level architecture with a meta-controller and a controller. The meta-controller provides the target performance in terms of KPIs, while the controller performs xApp orchestration at the lower level. Our simulation results show that the proposed HRL-based intent-driven xApp orchestration mechanism achieves $7.5\%$ and $21.4\%$ increase in average system throughput with respect to two baselines, i.e. a single xApp baseline and a non-machine learning-based algorithm, respectively. Similarly, $17.3\%$ and $37.9\%$ increase in energy efficiency is observed in comparison to the same baselines.
\end{abstract}

\begin{IEEEkeywords}
O-RAN, rApps, xApp, hierarchical reinforcement learning, orchestration
\end{IEEEkeywords}

\section{Introduction}
\label{s1}

Open radio access network (O-RAN) facilitates openness and intelligence to support diverse traffic types and their requirements in 5G and beyond networks\cite{7} as well as, multi-vendor RAN deployments. In a multi-vendor environment, rApps and xApps can be hosted in a non-real-time RAN intelligent controller (non-RT-RIC) and near-real-time RAN intelligent controller (near-RT-RIC). In the literature, xApps and rApps have been studied for resource and power allocation, beamforming and management, cell sleeping, traffic steering and so on\cite{2,3,4}. Advanced reinforcement learning (RL) algorithms can be used to develop intelligent network functions in O-RAN. However, a multi-rApp or a multi-xApp scenario with a variety of AI-enabled Apps will require intelligent control and orchestration among the Apps to avoid performance degradation.

Note that, we focus on xApps as a case study but our work generalizes to rApps as well. To elevate autonomy in O-RAN via xApp orchestration, intent-driven network optimization goals can play a pivotal role. The intent is defined as an optimization goal that is a high-level command given by the operator usually in plain language and it determines a key performance indicator (KPI) that the network should meet, such as ``increase throughput by $10\%$'' or ``increase energy efficiency by $5\%$'' \cite{23}. To better support autonomous orchestration of the xApps in a multi-vendor environment, emphasis on operators' intents is crucial \cite{24}. Intents aid in achieving agile, flexible, and simplified configuration of the wireless networks with minimum possible intervention. Furthermore, intelligent intent-driven management has the ability to constantly acquire knowledge and adjust to changing network conditions by utilizing extensive real-time network data. The inclusion of intent-driven goals for intelligent xApp control and orchestration is a promising yet highly complex task, since there are multiple vendors involved with different network functions and intents may trigger conflicting optimization goals in sub-systems. There are a few works on conflict mitigation or xApp cohabitation in O-RAN. For instance, Han \textit{et al.} propose a conflict mitigation scheme among multiple xApps using team learning \cite{8}, and Polese \textit{et al.} propose a machine learning (ML)-based pipeline for the cohabitation of multiple xApps in an O-RAN environment \cite {9}. The work outlined in \cite{26} introduces a method for achieving automation throughout the entire life cycle of xApps, beginning with the utilization scenario, requirements, design, verification, and ultimately, the deployment within networks. However, the operator intent is not involved in these works.  

To this end, we propose a hierarchical reinforcement learning (HRL) method for intent-driven xApp orchestration. Different from the previous works, the proposed scheme has a bi-level architecture, where we can pass the intents to the top-level hierarchy, and process it as optimization goals for the lower-level controller to control and orchestrate xApps. Orchestration can avoid xApp conflicts and improve performance by combining xApps with similar performance objectives. The proposed method is compared with two baselines: non-machine learning (non-ML) solution and a single xApp scenario. Our simulation results show that the proposed HRL-based intent-driven xApp orchestration mechanism achieves $7.5\%$ and $21.4\%$ increase in average system throughput along with $17.3\%$ and $37.9\%$ increase in energy efficiency, compared to the single xApp and non-ML baselines, respectively.

The rest of the paper is organized as follows: Section \ref{s2} discusses the related works, followed by Section \ref{s3} which presents the system model elaborately. The proposed HRL-based xApp orchestration in O-RAN is covered in Section \ref{s4}. Performance analysis and comparison of the proposed method along with the baselines are presented in Section \ref{s5}. Lastly, we present our conclusions in Section \ref{s6}.

\section{Related work}
\label{s2}
There are a few works that investigate ML-based xApps for RAN optimization and control. Polese \textit{et al.} propose an ML pipeline for multiple xApps in an O-RAN environment \cite{9}. Han \textit{et al.} propose a conflict mitigation scheme among deep reinforcement learning (DRL)-based power allocation and resource allocation xApps \cite{8}. Polese \textit{et al.} propose an Orchest-RAN scheme, in which network operators can specify high-level control objectives in non-RT-RIC to sort out the optimal set of data-driven algorithms to fulfill the provided intent \cite{11}. While the work presented in \cite {11} focuses on selecting the appropriate machine learning models and their execution locations for given inputs from the operator, it does not put emphasis on the network operator's goals as optimization objectives to select and orchestrate xApps.

An intent-driven orchestration of cognitive autonomous networks of RAN management is presented in \cite{12}, where the authors propose a generic design of intent-based management for controlling RAN parameters and KPIs. Zhang \textit{et al.} propose an intent conflict resolution scheme to realize conflict avoidance in machine learning-based xApps \cite{13}. A graph-based solution is proposed in \cite{14} to determine the specific network function required to fulfill an intent.  

Compared with existing literature, the main contribution of this work is that we propose an HRL scheme for intent-driven orchestration of xApps. The HRL scheme can well fit the inherent O-RAN hierarchy with non-RT-RIC and near-RT-RIC, and intent-based orchestration enables higher flexibility for network control and management. The intents from the human level operator are provided as goals for the system to achieve, which leads to the orchestration of xApps to achieve the provided goal.

\section{System Model}
\label{s3}
\subsection{System Model}
\begin{figure}[!t]
\centerline{\includegraphics[width=0.80\linewidth]{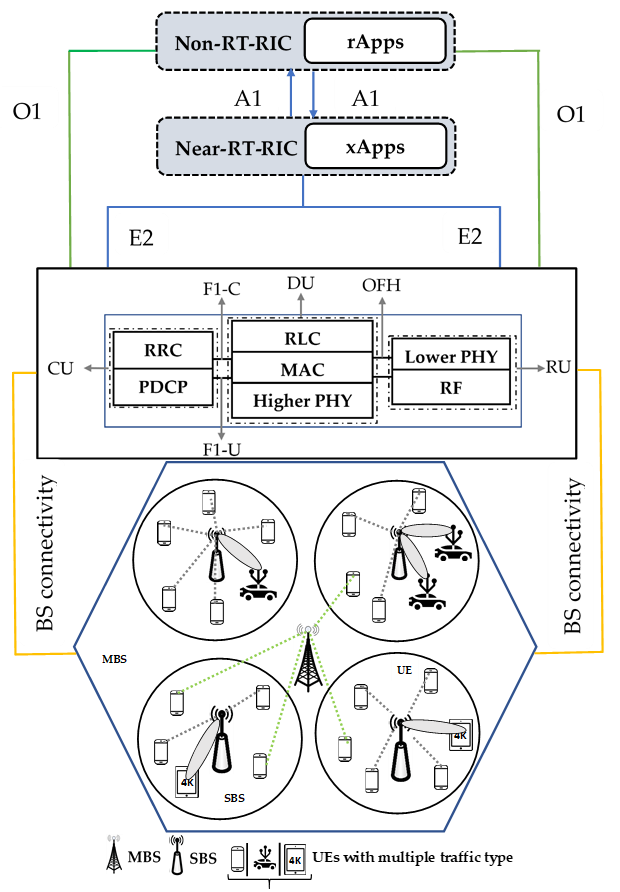}}
\caption{O-RAN based system model with macro cell and small cells.}
\label{fig1}
\vspace{-1.2em}
\end{figure}

We consider an O-RAN-based downlink orthogonal frequency division multiplexing cellular system having $B$ BSs serving $U$ users simultaneously, and multiple small cells in the system are within the range of a macro cell. There are $K$ classes of traffic in the system and users are connected with multiple RATs via dual connectivity. There are $Q$ classes of RATs ($q_1,q_2,...,q_n$), where $q$ represents a certain access technology (LTE, 5G, etc.). The wireless system model considered in this work is presented in Fig. \ref{fig1}. RIC platforms in the figure (non and near-RT-RIC) can host rApps and xApps which are control and optimization applications operating at different time scales.

We design three xApps, namely traffic steering, cell sleeping, and beam forming xApps. In each xApp, we apply deep reinforcement learning for optimization within this xApp, which will be introduced in the following.

\subsubsection{Traffic Steering xApp}

The traffic steering xApp aims to achieve a simultaneous balance of QoS requirements for various traffic classes by introducing a traffic steering scheme based on Deep Q-Network (DQN) \cite{17}. We design the reward and state functions to ensure satisfactory performance, focusing on two essential KPIs: network delay and average system throughput. Traffic can be steered to a certain BS based on load experienced, link quality, and traffic type. The details of this xApp can be found in \cite{2}.

\subsubsection{Cell Sleeping xApp}

The cell sleeping xApp is designed to reduce power consumption in the system by turning off idle or less busy BSs. The xApp can perform cell sleeping based on traffic load ratios and queue length of each BS. The energy consumption model for the BS is:
\begin{equation} \label{eq2}
 P_{in}=\left\{
\begin{array}{lcl} P_{0}+\delta_{p} P_{out},     &      & 0 < P_{out} \leq P_{max}, \\
P_{sleep} ,    &      &  P_{out}=0,
\end{array} \right.
\end{equation}
where $P_{0}$ is the fixed power consumption, $\delta_{p}$ is the slope of load-dependent power consumption, $P_{out}$ is the transmission power, $P_{max}$ is the maximum transmission power, and $P_{sleep}$ is the constant power consumption in sleep mode \cite{16}. 

The goal of the cell sleeping xApp is to maximize energy efficiency as much as possible without overloading the active BSs. The optimization goal is formulated as follows:
\begin{equation} \label{eq3}
    \begin{split}
     \max_{P_b} \quad \frac{\sum_{u\in {U_o}}\sum_{b\in B} T_{u,b}}{P_b}-\theta b_u, \\
     s.t. \quad \quad \quad \quad(\ref{eq2}), \quad\quad\quad\quad\quad\quad\quad\\  
    \end{split}
\end{equation}
where $U_o$ is the set of the user equipments (UEs) connected to a certain BS, $T$ represents the throughput, $\theta$ is the penalty factor to reduce overloading, and $b_u$ is the number of the BSs overloaded. Turning off the BSs can greatly decrease power consumption. It reduces the number of BSs active that are serving the live network traffic. This poses a risk of overloading the active BSs. Therefore, the penalty factor related to the number of BSs has been introduced to avoid excessive overloading. 

To address the formulated problem, the following MDP is formulated: 
\begin{itemize}
\item \textbf{State:} The set of the state consists of: $S=\{q_L, L_R\}$. $L_R$ represents the traffic load ratio of a BS, $b$. The second element of the state space is the queue length of the BSs representing the load level.  
\item \textbf{Action:} Turning the BSs on and off are put into the action set for the DQN implementation. $A=\{ON, OFF\}$. 
\item \textbf{Reward:} The reward function is the same as eq. (\ref{eq3}). 
\end{itemize}

\subsubsection{Beamforming xApp}

The third xApp is the beamforming xApp.         
We deploy band-switching BSs from 3.5 GHz to mmWave frequencies \cite{18}. This allows us to support high throughput traffic like enhanced mobile broadband (eMBB) via accurate intelligent beamforming. This xApp can control power based on the location of the UE and it uses minimum transmission power needed which is energy efficient. The xApp employs analog beamforming, and a multi-antenna setup is adopted where each BS deploys a uniform linear array (ULA) of $M$ antennas \cite{25}. The beamforming weights of every beamforming vector are implemented using constant modulus phase shifters. We also assume that there is a beam steering-based codebook, $F$, from which every beamforming vector is selected \cite{25}.

Every BS $l$ has a transmit power $P_{TX,l} \in P$, where $P$ is the set of candidate transmit powers. We want to optimize two parameters: throughput and energy efficiency using this xApp. To obtain such a goal, the following optimization problem is addressed.  
\begin{equation}
     \begin{split} \label{eq4}
      max\sum_{l\in \{1,2,..,L\}}\left[c_1\left(\frac{T_{k,b}}{T_{QoS}}\right) + c_2\left(\frac{\varepsilon}{\varepsilon_{max}}\right)\right], \\
     s.t. \quad \quad \quad \quad \quad \quad  P_{TX,l}[t] \in P, \quad \quad \quad \quad \quad \quad\\
     f_l[t]\in F, \quad \quad \quad \quad \quad \quad \quad
     \end{split}
\end{equation}
where $T_{k,b}$ is the throughput achieved by the system, $T_{QoS}$ is the defined throughput requirement for a traffic type $k$, $\varepsilon$ represents the energy efficiency associated with the BS throughput and transmission power, $\varepsilon_{max}$ is the maximum theoretical energy efficiency, and $c_1$ and $c_2$ are the weight factors.

To solve the formulated problem, the following MDP is defined. 
\begin{itemize}
\item \textbf{State:} UE co-ordinates are used as set of states, $S=\{C_{UE1}, C_{UE2},...,C_{UEN}\}$.  
\item \textbf{Action:} The action set consists of two elements: $A=\{\alpha(\chi_n),\delta_n\}$. Here, $\chi$ is the steering angle, and $\alpha(\chi_n)$ is the array steering vector in the direction $\chi_n$ of the $n$-th element in the codebook. $\delta_n$ accounts for the power level change. 
\item \textbf{Reward:} The reward function is the same as eq. (\ref{eq4}) as presented before. 
\end{itemize}

\section{Proposed HRL-based xApp orchestration Scheme}
\label{s4}
RL problems can be formulated as MDPs where we have a set of states, actions, transition probability, and a reward function ($S,A,T,R$). The RL agent in HRL consists of two controllers: a meta-controller and a controller \cite{10}. The MDP for HRL has an added element which is denoted as a set of goals ($G$). Depending on the current state, the meta-controller is responsible for generating high-level goals ($G={g_1,g_2,...,g_n}$) for the controller. After that, these goals are transformed into high-level policies. The controller chooses low-level action $a$ according to the high-level policies. This process from the controller yields an intrinsic reward ($r_{in}$). Finally, an extrinsic reward ($r_{ex}$) is given to the meta-controller from the environment and it will provide the controller with a new goal ($g_\prime$). This section will discuss the xApp orchestration scheme via HRL.     

\subsection{xApp Coordination Using HRL}

\begin{figure}[!t]
\centerline{\includegraphics[width=0.80\linewidth]{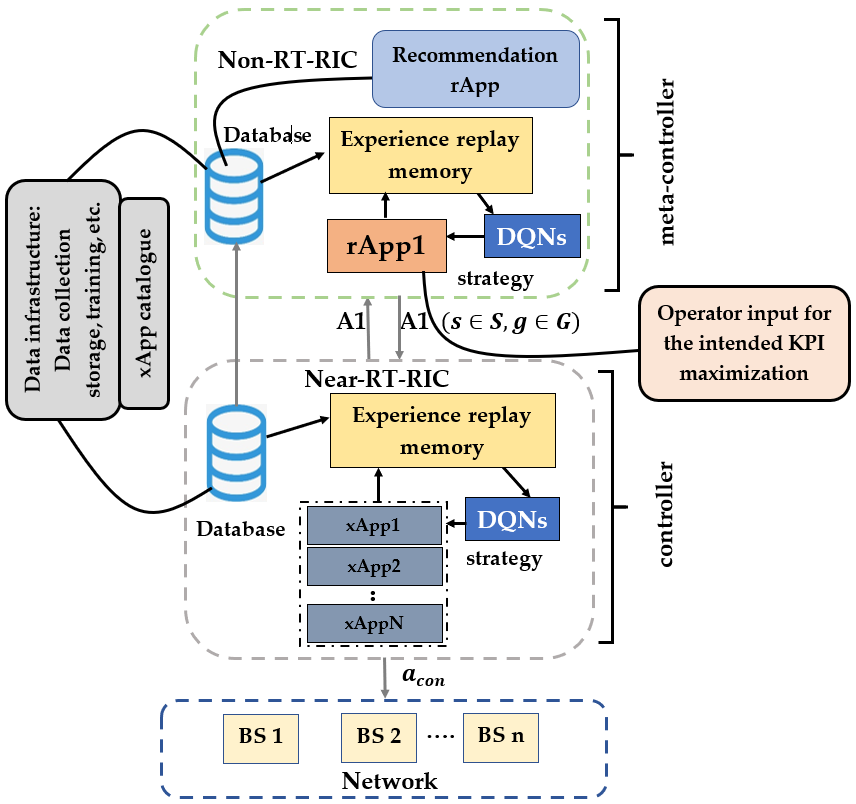}}
\caption{Intent-based xApp orchestration with macro and micro cells having different types of traffic.}
\label{fig2}
\vspace{-1.2em}
\end{figure}

The proposed O-RAN-based system architecture is presented in Fig \ref{fig2}. RIC platforms can host rApps and xApps which are applications operating at different time scales. Three xApps have been defined in previous sections. The rApp in the figure works as an input panel for the network operator, and it can convert these inputs as goals to be optimized. Also, it works as the meta-controller in the non-RT-RIC.

Let's assume, $X$ is a set of xApps and $Y$ is the subset of $X$ having at least one element (xApp in our case), that can optimize the network performance based on the operator input. Let $I$ be the set of candidate KPIs that a xApp can optimize and $Z$ be the set of QoS requirements the system has to satisfy. Considering all these assumptions, the xApp orchestration problem that we want to address can be formulated as follows: 

\begin{equation} \label{eq5}
\begin{split}
     max\sum_{i\in I}\sum_{z\in Z}(P_i-\rho \xi_z), \quad\quad\quad  \\
     s.t. \quad \forall(X)\exists(Z): V(O)=1, \quad\quad
\end{split}
\end{equation}
where $P$ is the intended performance metric the operator intends to improve. $\rho$ is the penalty parameter for QoS requirement violation, and $\xi_z$ is the number of UEs QoS requirements violated to. Lastly, $V(O)$ is the proposition that ``An xApp can improve a performance metric", which is either ‘0’ or ‘1’. 

As presented in Fig. \ref{fig2}, the rApp in the system is directly connected to the user panel where the operator may provide input to the system. The operator input is provided as the percentage of the increase related to a certain KPI. For example, $x\%$ for throughput increase or $y\%$ for energy efficiency increase or any other intent stated in natural language. The rApp has a hierarchical deep Q-learning (h-DQN) framework \cite{10}. The meta-controller (in non-RT-RIC) takes the increased amount of throughput or increased amount of energy efficiency as a goal, observes the state in the environment and provides both the goal and states to the controller in near-RT-RIC having a bundle of xApps. This type of data passing is done via the A1 interface by which both the non and near-RT-RIC are connected. The controller takes the action of choosing an xApp or a set of xApps based on the provided state and goal. Following, we define the MDP for the meta-controller and controller to address the xApp orchestration problem formulated in eq. (\ref{eq5}).
\begin{itemize}
\item \textbf{State:} The set of states consists of traffic flow types of different users in the network. UEs having similar traffic types are grouped together. $S=\{T_{voice},..,T_{urllc},...,T_{gaming},...,T_{eMBB},..\}$. Elements in this set stand for five different traffic types in the system. Both meta-controller and controller share the same states.
\item \textbf{Action:} xApp or combination of xApp selection is considered as actions to be performed by the controller which is defined as: \{$A_{xApp1}, A_{xApp1,2},...., A_{xAppN}\}$.
\item \textbf{Intrinsic reward:} The intrinsic reward function ($r_{in}$) for the controller is: $r_{in}=P_i-\rho \xi_z$ which is similar to eq. (\ref{eq5}). 
\item \textbf{Goal for the controller:} Increased throughput or increased energy efficiency level that can satisfy operator intent is passed to the controller as goals. It is $G=\{tp_1, tp_2,..., tp_n\}$ for throughput increasing intents or $G=\{ee_1, ee_2,..., ee_n\}$ for energy efficiency increasing intents. Note that these goals can be generalized to other KPIs however for simplicity we target throughput and energy efficiency.    

\item \textbf{Extrinsic reward:} The meta-controller is responsible for the overall performance of the system. Therefore, we have set the extrinsic reward function for the meta-controller as the objective of the problem formulation presented in eq. (\ref{eq5}). The following equation is basically the summation of the intrinsic reward over $\tau$ steps. 
\begin{equation}
r_{ex,\tau}= \frac{1}{n}\sum_{\tau=1}^{n} r_{in,\tau} \quad 
\forall(u)\in U,\forall(b)\in B,
\end{equation}
\end{itemize}

The whole process of xApp orchestration can be summarized as follows: 
\begin{itemize}
\item \textbf{Step 1:} Operator's intent is provided as input regarding which performance metric is to be improved. 
\item \textbf{Step 2:} These performance targets are provided to the controller in near-RT-RIC by the meta controller rApp in the non-RT-RIC as goals to achieve.
\item \textbf{Step 3:} The controller selects an xApp or a combination of xApps to reach the target performance as close as possible. The system learns based on the reward it gets for such kind of xApp selection. 
\item \textbf{Step 4:} Selected xApps with their own DRL-based functionalities optimize the performance of the network as a response to the intent of the operator. 
\end{itemize}

\subsection{Baseline Algorithms}

This section includes two baselines. The first baseline is a simulation of the same network scenario based on the system model we have presented so far where there is no intelligent DRL-based xApp to optimize the network. We use non-ML algorithms. For comparing the throughput performance of the proposed HRL-based system, we use a threshold-based traffic steering scheme proposed in \cite{27}.  It uses a predefined threshold. The threshold is determined considering the load at each station, channel condition, and user service type. The mean of all these metrics is taken to obtain the threshold ($T$) values. Weighted summation of the same parameters is taken to form a variable ($w$). Then, the traffic is steered to another BS based on the $w$ and $T$ values. This baseline does not include cell sleeping, therefore BSs are always on. In our second baseline, we consider single xApp scenarios. For example, the proposed HRL-based xApp orchestration mechanism is compared with single xApp scenarios where only traffic steering xApp is in action, or the cell sleeping xApp is in action.

\section{Performance Evaluation}
\label{s5}
\subsection{Simulation setup}
A MATLAB-based simulation environment has been developed having one eNB and four gNBs to serve as one macro-cell and four small cells. In total, we deploy 60 UEs with five different traffic types: voice, gaming, video, URLLC, and eMBB. Different types of traffic in the system have variant requirements in terms of different KPIs. QoS requirements of different traffic types have been defined based on our previous work \cite{22}. eMBB and URLLC traffic types have been added additionally to test the system compatibility. For the eMBB traffic type, we consider packet size, $T_{QoS}$, and $D_{QoS}$ to be 1500 bytes, 100 Mbps, and 15 ms, respectively \cite{30}. Lastly, specifications related to delay and packet size for the URLLC traffic are set to 32 bytes and 2.5 ms. 

A 5G NSA mode having different types of RAT (LTE and 5G) in the simulation environment work together. We deploy an architecture based on \cite{21}. The carrier frequency for LTE is set to be 800 MHz. For 5G NR small cells, band-switching BSs are deployed at 3.5 GHz and 30 GHz. BS transmission power for LTE and 5G NR is set to be 38 dBm and 43 dBm, respectively \cite{28}. 

For the HRL implementation, the starting rate of learning is set to 0.95. In order to maintain stable learning performance, we reduce the learning rate periodically after a certain number of episodes. Additionally, the discount factor used is 0.3. The simulation is conducted 10 times using MATLAB, and the average outcomes are presented along with a 95\% confidence interval.

\subsection{Simulation results}

\begin{figure}[!t]
\centerline{\includegraphics[width=0.92\linewidth]{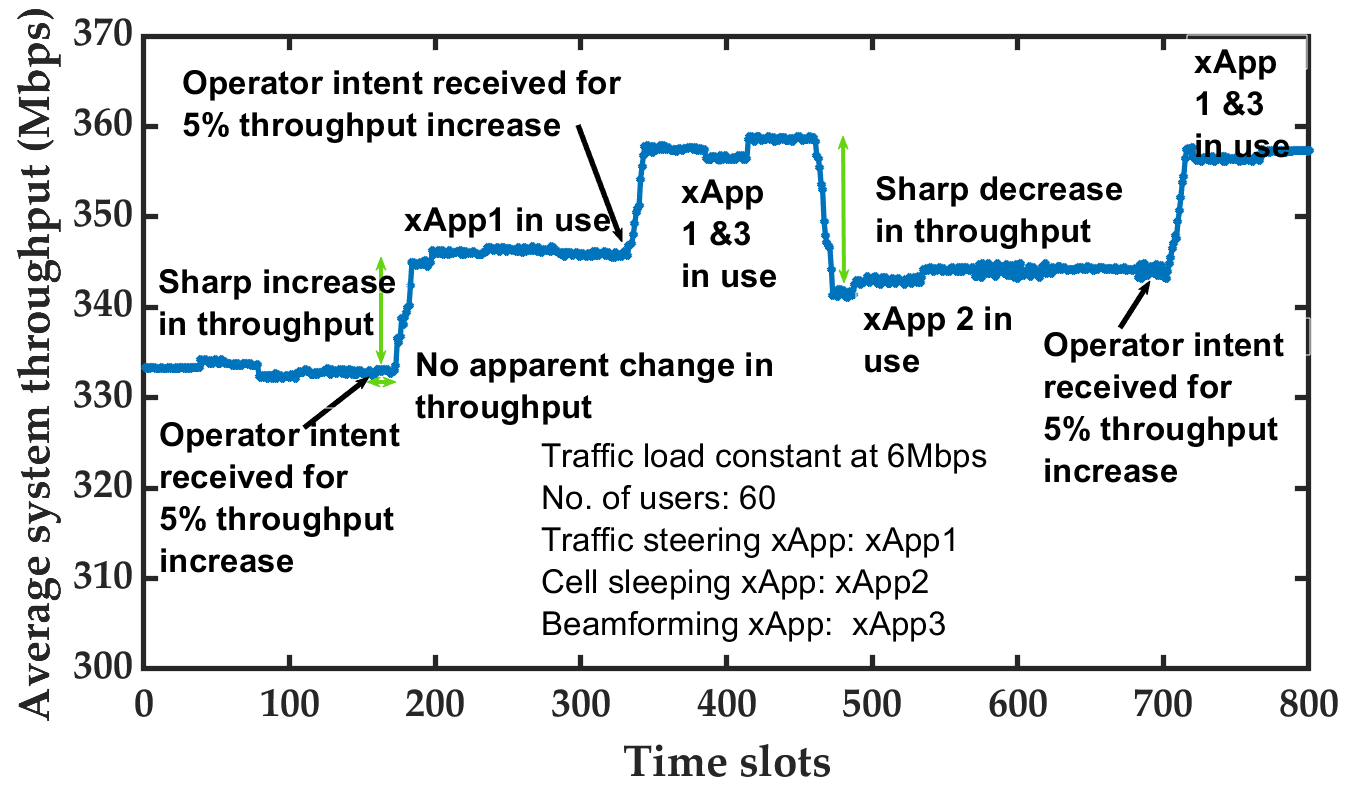}}
\caption{Impact of operator intents on throughput.}
\label{fig3}
\vspace{-1.2em}
\end{figure}

\begin{figure}[!t]
\centerline{\includegraphics[width=0.92\linewidth]{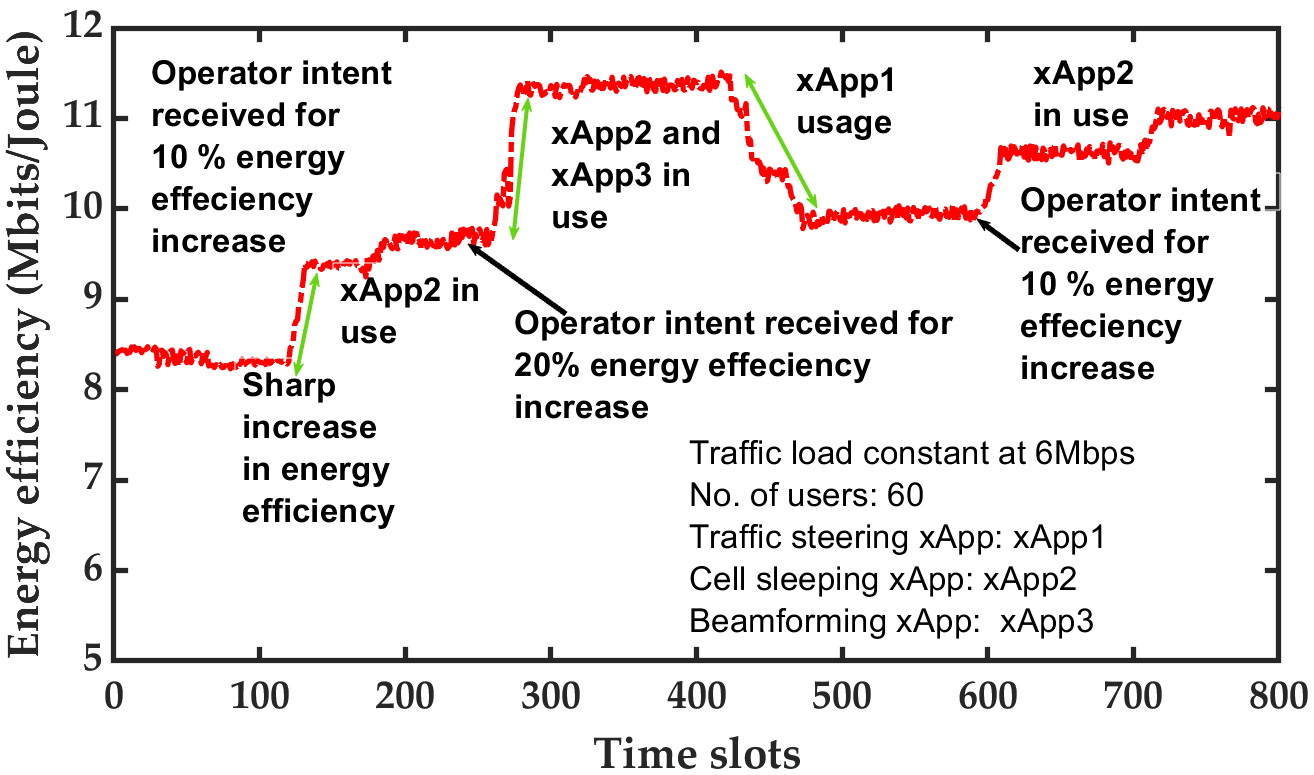}}
\caption{Impact of operator intents on energy efficiency.}
\label{fig4}
\vspace{-1.2em}
\end{figure}

Before conducting the performance evaluation of the proposed xApp orchestration scheme, first, we present how the intent-oriented HRL-based orchestration scheme works. Fig. \ref{fig3} shows that the operator intent of ``increase throughput'' leads to the selection of certain xApps. When there is a $5\%$ throughput increase intent from the operator, after a few time slots, there is a sharp increase in throughput. This is because xApp1 (traffic steering xApp) has been invoked. When a 5\% increase is again given as an input, a combination of xApp1 and xApp3 (intelligent beamforming xApp) is selected. When the operator provides an intent to decrease power consumption by $5\%$, we can see from Fig. \ref{fig3} that there is a sharp decrease in throughput. This is because xApp1 and xApp3 have been terminated at the 461-th time slot and xApp2 (cell sleeping xApp) has been invoked.   

Fig. \ref{fig4} presents a similar graph to the previous one but this time it plots the energy efficiency in the time axis. When there is an intent from the operator to achieve ``$10\%$ increase in energy efficiency'', we can see that there is an initiation of xApp2 at the 131-th time slot. This xApp performs cell sleeping and saves energy. For the next energy efficiency increase intent given by the operator, it can be seen that both xApp2 and xApp3 are working together. The proposed HRL-based algorithm has successfully orchestrated these two xApps for the desired performance gain. Fig. \ref{fig3} and \ref{fig4} basically show the utility of the proposed system. Not only it can induce operator intent as an optimization goal, but also it can orchestrate xApps to gain desired performance output by using the proper combination of xApps.

\begin{figure}[!t]
\centerline{\includegraphics[width=0.80\linewidth]{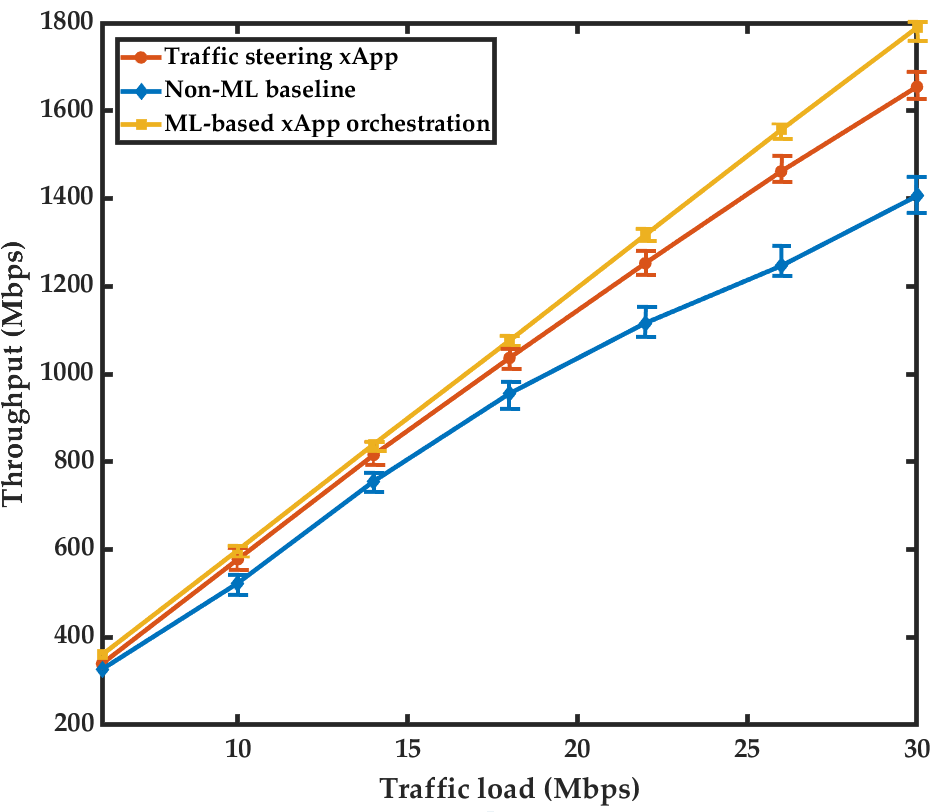}}
\caption{Performance comparison in terms of average system throughput.} 
\label{fig5}
\vspace{-1.2em}
\end{figure}

Fig. \ref{fig5} shows the performance comparison between the proposed HRL-based xApp orchestration scheme and the baseline scenarios in terms of average system throughput. Results are obtained under a constant load of 6 Mbps. The proposed orchestration scheme achieves a $21.4\%$ increase and $7.5\%$ increase in average system throughput compared to the non-ML algorithm and single xApp scenario (traffic steering xApp), respectively. It is because of the efficient orchestration mechanism that involves multiple xApps that trigger the optimal combination of xApps to reach better performance based on the operator intent.  

\begin{figure}[!t]
\centerline{\includegraphics[width=0.80\linewidth]{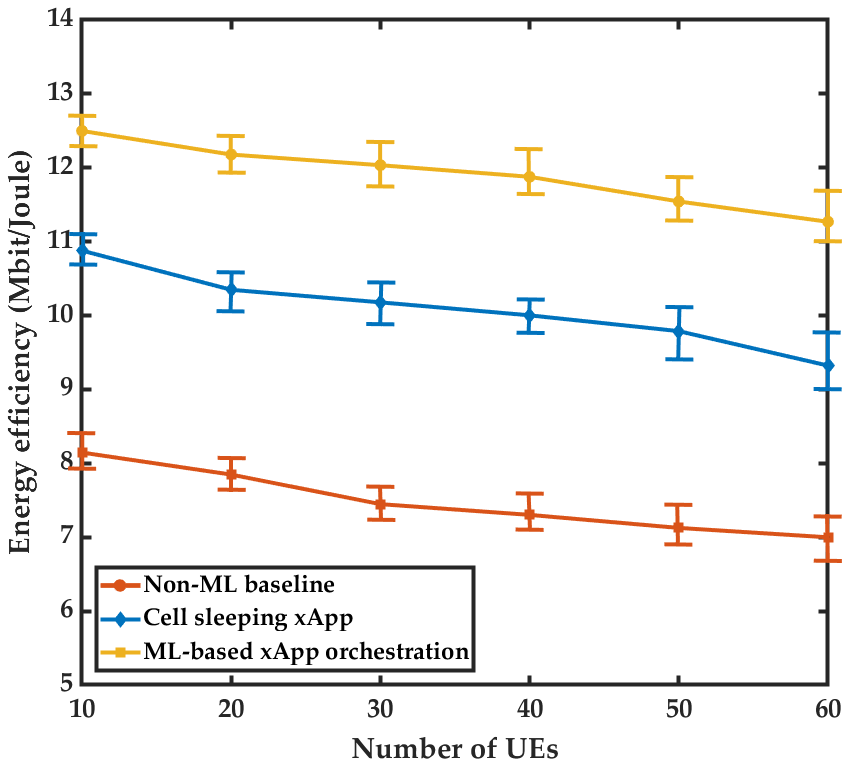}}
\caption{Performance comparison in terms of energy efficiency.} 
\label{fig6}
\vspace{-1em}
\end{figure}

Fig. \ref{fig6} shows the performance comparison between the proposed HRL-based xApp orchestration scheme and the baseline scenarios in terms of average energy efficiency. The proposed orchestration scheme obtains a $17.3\%$ increase and $37.9\%$ increase in average energy efficiency compared to the single xApp and non-ML scenario (cell sleeping xApp), respectively. Similar to the former, it is because of the HRL-based orchestration mechanism that incorporates multiple xApps to achieve better performance based on the user intent. Also, note that we use traffic steering in the former figure and cell sleeping in this evaluation because they are specifically optimizing throughput and energy respectively.      

\section{Conclusions}
\label{s6}
In this paper, we show that the HRL-based intent-driven orchestration mechanism is vastly effective in not only optimizing KPIs but also providing great flexibility and control to the operator. In this study, we have introduced a novel HRL-based xApp orchestration mechanism that can perform xApp management and provide recommendations for the best combination of xApps given the operator's intent. The optimal xApp orchestration scheme has led to a $7.5\%$ increase in average system throughput and a $17.3\%$ increase in energy efficiency compared to single xApp usage with no orchestration. In our future work, we plan to extend this orchestration to rApps and other xApps with complex KPI interactions.

\section*{Acknowledgement}
This work has been supported by MITACS and Ericsson Canada, and NSERC Canada Research Chairs and NSERC Collaborative Research and Training Experience Program (CREATE) under Grant 497981.

\bibliographystyle{IEEEtran}
\bibliography{reference.bib}
\end{document}